\documentclass[manuscript]{acmart}
\newcommand{\icon}[1]{\noindent
\raisebox{-2.5pt}{\setlength{\fboxsep}{0pt} \setlength{\fboxrule}{0pt}{\includegraphics[height=\baselineskip]{#1}}}}

\AtBeginDocument{%
  \providecommand\BibTeX{{%
    \normalfont B\kern-0.5em{\scshape i\kern-0.25em b}\kern-0.8em\TeX}}}

\copyrightyear{2023}
\acmYear{2023}
\setcopyright{acmlicensed}\acmConference[CHI '23]{Proceedings of the 2023 CHI Conference on Human Factors in Computing Systems}{April 23--28, 2023}{Hamburg, Germany}
\acmBooktitle{Proceedings of the 2023 CHI Conference on Human Factors in Computing Systems (CHI '23), April 23--28, 2023, Hamburg, Germany}
\acmPrice{15.00}
\acmDOI{10.1145/3544548.3581247}
\acmISBN{978-1-4503-9421-5/23/04}

\usepackage{colortbl}
\usepackage{enumitem}
\usepackage{multirow}
\usepackage{csquotes}

\def\markup{0}
\if\markup1
\newcommand{\rv}[1]{{\leavevmode\color{blue}#1}}
\else
\newcommand{\rv}[1]{#1}
\fi

\def\minor{1}
\if\minor1

\else

\fi

\begin{document}



\title[]{Collaboration with Conversational AI Assistants for UX Evaluation: Questions and How to Ask them (Voice vs. Text)}




\author{Emily Kuang}
\affiliation{%
  \institution{Golisano College of Computing and Information Sciences}
  \institution{Rochester Institute of Technology}
  \city{Rochester}
  \state{New York}
  \country{USA}
}
\email{ek8093@rit.edu}

\author{Ehsan Jahangirzadeh Soure}
\affiliation{%
  \institution{School of Computer Science}
  \institution{University of Waterloo}
  \city{Waterloo}
  \state{Ontario}
  \country{Canada}
}
\email{ejahangi@uwaterloo.ca}

\author{Mingming Fan}
\orcid{0000-0002-0356-4712}
\authornote{Corresponding Author}
\affiliation{
  \institution{Computational Media and Arts Thrust}
  \institution{The Hong Kong University of Science and Technology (Guangzhou)}
  \city{Guangzhou}
  \country{China}
}
\affiliation{
  \institution{Division of Integrative Systems and Design \& Department of Computer Science and Engineering}
  \institution{The Hong Kong University of Science and Technology}
  \city{Hong Kong SAR}
  \country{China}
}
\email{mingmingfan@ust.hk}

\author{Jian Zhao}
\affiliation{%
  \institution{School of Computer Science}
  \institution{University of Waterloo}
  \city{Waterloo}
  \state{Ontario}
  \country{Canada}
}
\email{jianzhao@uwaterloo.ca}

\author{Kristen Shinohara}
\authornotemark[1]
\affiliation{%
  \institution{School of Information}
  \institution{Rochester Institute of Technology}
  \city{Rochester}
  \state{New York}
  \country{USA}
}
\email{kristen.shinohara@rit.edu}

\renewcommand{\shortauthors}{Emily Kuang, et al.}

\begin{abstract} 
AI is promising in assisting UX evaluators with analyzing usability tests, but its judgments are typically presented as non-interactive visualizations. 
Evaluators may have questions about test recordings, but have no way of asking them.
Interactive conversational assistants \rv{provide a Q\&A dynamic that may improve analysis efficiency and evaluator autonomy.} 
\rv{To understand the full range of analysis-related questions, we conducted a Wizard-of-Oz design probe study with 20 participants who interacted with simulated AI assistants via text or voice.}
We found that participants asked for \rv{five} categories of information: user actions\rv{, user} mental model, help from the AI assistant, product and task information, and user demographics.
Those who used the text assistant asked more questions, but the question lengths were similar.
The text assistant was perceived as significantly more efficient, but both were rated equally in satisfaction and trust. 
We also provide design considerations for future conversational AI assistants for UX evaluation.
\end{abstract}





\maketitle

\section{Introduction}

Usability testing is a frequently employed user-centered design method for detecting usability problems, but analyzing test recordings is tedious, challenging, and time-consuming~\cite{norgaard_what_2006, chilana_understanding_2010, folstad_analysis_2012, fan_practices_2020}. 
UX evaluators must take into account behavioral signals in both the visual and audio channels of usability test recordings while assessing multiple tasks simultaneously at a fast pace~\cite{chilana_understanding_2010}.
In industry, UX evaluators also have limited time and resources, which could lead to missed information or misinterpreted problems \cite{hertzum_evaluator_2001, norgaard_what_2006, folstad_analysis_2012, kuang_merging_2022}.
Despite the value of working with others to improve reliability and completeness, few evaluators employ collaboration in practice \cite{folstad_analysis_2010, folstad_analysis_2012, kuang_merging_2022}. 
In an international survey of 279 UX evaluators, only 37\% reported collaboration when analyzing the same recordings \cite{kuang_merging_2022} and in other cases, it was found that matching teams or pairs were costly in terms of time, resources, and effort \cite{fan_practices_2020}.
Thus, individual analysis \rv{of usability recordings} can be problematic, but effective collaboration is often hindered by limited resources.

To address the shortage of human-human collaboration, AI-driven analysis to aid UX evaluators is considered an effective tool, particularly for common usability issues, that could boost the efficiency of UX evaluators and the reliability of results \cite{kuang_merging_2022}.
Some commercial analytical platforms already contain features derived from AI and machine learning (ML) (e.g., UserTesting offers sentiment analysis \cite{usertesting_usertesting_2021}, and UXTesting offers emotion detection \cite{uxtesting_uxtesting_2022}). 
Researchers have also incorporated ML and AI into the UX field \cite{paterno2017customizable,harms2019automated,jeong2020detecting, yang_re-examining_2020, zimmerman_ux_2020, oztekin_machine_2013} and created algorithms that were shown to detect similar usability problems as manual testing \cite{grigera2017automatic}.
However, such automated methods did not thoroughly identify problems and could not replace the human reasoning required for application-specific problems \cite{grigera2017automatic}.
To address these limitations, recent work developed human-AI collaborative tools where UX evaluators can utilize visualizations of ML-driven features to inform their UX analysis \cite{fan_vista_2020, soure_coux_2021, fan_human-ai_2022}.
\rv{Yet, these non-interactive visualizations provided information regardless of whether they were needed by UX evaluators and fell short in addressing specific questions that UX evaluators may have about observations from the recordings \cite{soure_coux_2021}.}
There is increasing evidence that domain experts preferred to treat ML and AI models as ``another colleague'' and receive \rv{information} in the form of natural language dialogues \cite{lakkaraju_rethinking_2022}. 
Thus, we consider that an interactive assistant---in the form of a conversational agent---may provide an opportunity for a Q\&A dynamic that \rv{presents information on demand and improves analytic efficiency}.

Conversation is becoming a key mode of human-computer interaction due to the proliferation of conversational agents \cite{luger_like_2016}.
Conversational agents are increasing in both professional and personal use, where 70\% of white-collar workers are expected to interact with text chatbots on a daily basis in 2022 \cite{goasduff_chatbots_2019}, and over half (56.4\%) of smartphones owners utilized the built-in voice assistant in 2020 \cite{kinsella_voice_2020}.
Since text and speech are the two main ways to interact with conversational agents, prior work has compared the two modalities and demonstrated strong differences in user behavior between them \cite{kocielnik_designing_2018, luria_comparing_2017, neuwirth_distributed_1994, kang_understanding_2017}.
For example, text agents designed for journaling and reflection were considered more familiar and efficient than voice ones \cite{kocielnik_designing_2018}.
On the other hand, voice comments were more positively received than those in text for collaborative writing tasks \cite{neuwirth_distributed_1994}.
Since the context of use impacts user preferences for either text or voice interactions, we seek to understand the differences between these modalities within the context of UX evaluation.

\rv{In this research, we take the first step toward determining the expected functionalities and desired interactions with a conversational assistant for UX analysis.}
To build AI assistants that could respond to a full range of questions \rv{about usability test recordings} from UX evaluators, we must first understand what that full range might be.
Thus, we conducted a design probe in which evaluators used a \rv{simulated} AI assistant with two modalities (voice and text) to ask any questions that they considered relevant to their analysis. 
Using the \rv{simulated} AI assistant as a probe, we investigated if the Q\&A dynamic and modality of interaction \rv{provided viable support to UX evaluators during analysis.} 
Specifically, our study was guided by the following research questions (RQs):
\begin{itemize}
    \item \textbf{RQ1} - What types of questions will UX evaluators ask an AI assistant during analysis?
    \item \textbf{RQ2} - How do the number and content of questions asked by UX evaluators differ between text and voice interactions?
    \item \textbf{RQ3} - What are the participants' perceptions of text and voice assistants?
\end{itemize}

To circumvent the technical limitations of developing AI algorithms that could consistently extract accurate information from usability test recordings, we adopted a Wizard-of-Oz approach.
During the study, participants asked questions \rv{about the usability test recordings} via text or voice to the AI assistant, which was simulated by the researchers. 
We then analyzed the study sessions by coding participants' questions and grouping them into categories, identifying differences between interaction modalities, and analyzing the post-study interview and survey responses.

The results indicate that participants were interested in \rv{five} categories of information: user actions, \rv{user} mental model, help from the AI assistant, product and task information, and user demographics.
Those who used the text assistant tended to ask more questions, but the lengths of the questions were not significantly different between conditions.
The text assistant was perceived as significantly more efficient, but both were rated equally in satisfaction and trust. 
Based on the questions and feedback that participants provided, we derived design considerations for future conversational AI assistants, \rv{which includes} consolidating analysis from multiple recordings and allowing evaluators to choose the modality of interaction.
In sum, we make the following contributions:
\begin{itemize}
\item We present a dataset of 325 questions from a design probe study to understand the types of questions UX evaluators are interested in asking conversational AI assistants \rv{about usability test recordings}; 
\item We show differences between text and voice interactions \rv{with a simulated AI assistant} for UX analysis; 
\item We highlight design considerations for improving future conversational AI assistants for UX analysis. 
\end{itemize}

\section{Background and Related Work}

Our work is informed by prior research on the importance of collaboration in UX analysis, machine learning for UX analysis, and human-AI collaboration via text and voice assistants. 

\subsection{Importance of Collaboration in UX Analysis}
The most frequently employed method for detecting usability problems with digital products is through usability testing \cite{fan_practices_2020}. 
UX evaluators assess both the visual and audio channels of recordings, observing user actions and writing notes simultaneously \cite{chilana_understanding_2010}. 
However, analyzing usability test recordings using these manual approaches is challenging and time-consuming because evaluators have limited time and resources, which could lead to missed information or misinterpreted problems \cite{hertzum_evaluator_2001, norgaard_what_2006, folstad_analysis_2012, fan_vista_2020}.
To balance analytic reliability and validity with efficiency, evaluators collaborate in pairs or teams \cite{fan_practices_2020, hertzum_evaluator_2001}. 
Collaborations include reviewing recordings together, and are shown to divide the workload and comprehensively detect problems \cite{hertzum_evaluator_2001, hertzum_what_2013, fan_practices_2020, kuang_merging_2022}.
Collaborations also alleviate the \textit{``evaluator effect,''} the condition in which different evaluators identify different sets of UX problems even when analyzing the same test session \cite{hertzum_evaluator_2001,jacobsen_evaluator_1998}, and therefore ensures comprehensive evaluative coverage. 
Thus, collaborations benefit from different perspectives, increasing reliability \cite{hertzum_evaluator_2001} and thoroughness of the problems identified \cite{sears_heuristic_1997}.

However, despite the value of working with others to improve reliability and completeness, few evaluators employed collaboration in practice since it was costly in terms of time, resources, and effort \cite{folstad_analysis_2012, kuang_merging_2022, fan_practices_2020}.
When UX evaluators did collaborate, over two-thirds of them found it difficult to coordinate and merge analysis from multiple collaborators \cite{kuang_merging_2022}.
Thus, individual analysis can be problematic, but effective collaboration is often stymied by limited resources. 
By contrast, AI-driven analysis to aid UX evaluators may be an effective tool that could reduce the overhead costs associated with collaboration while boosting the reliability of results. 
Advances in natural language processing and ML enable automatic cues detection from acoustic, textual, and visual channels available in recordings \cite{fan_automatic_2020, fan_vista_2020}. 
AI technology, harnessed appropriately, could assist UX evaluators with identifying usability problems by taking advantage of improvements in technology, systematically uncovering predictable aspects of usability analysis, and alleviating costs associated with collaboration \cite{fan_automatic_2020, grigera2017automatic}. We consider the different possibilities explored in the next section. 

\subsection{Machine Learning for UX Analysis}
Given the potential benefits of integrating AI in usability analysis, researchers examined how to use AI to detect UX problems \rv{by creating ML classifiers based on user interaction events} \cite{grigera2017automatic,paterno2017customizable,harms2019automated,jeong2020detecting, yang_re-examining_2020, zimmerman_ux_2020, oztekin_machine_2013}. 
However, automated algorithms were still unable to detect the full set of problems that were identified with manual analysis \cite{grigera2017automatic}. 
These results indicate that although automated methods can find meaningful problems, they cannot replace human reasoning required for completeness. 
Furthermore, these automatic methods were primarily based on users' interaction logs, which only reflect some aspects of UX problems, and which do so without direct observation of user behavior \cite{jeong2020detecting}.

Due to the limitations of automated methods, there is growing interest in human-AI collaboration where human decision-making is supplemented with AI assistance \cite{lai_towards_2021}.
Recent work developed tools where UX evaluators can utilize visualizations of ML-driven features to inform their identification of usability problems \cite{fan_vista_2020, soure_coux_2021}.
However, these tools offered non-interactive visualizations and participants in their study did not have a way to express their questions towards certain ML-driven features or ask for an explanation of the underlying algorithm \cite{soure_coux_2021}. 
Other work explored the effects of explanation and synchronization on UX analysis and found that AI with explanations provided better support and was perceived more positively \cite{fan_human-ai_2022}. 
However, all aforementioned works directly presented information extracted from usability test recordings to UX evaluators without taking into account what they found to be most valuable.
Based on the limitations of non-interactive visualizations and predetermined information, we consider that an interactive assistant---in the form of a conversational agent---may provide timely information on an as needed basis.

\subsection{Human-AI Collaboration via Interactive Conversational Assistants}

There is a growing trend in the usage of conversational agents in our daily lives, and conversation is becoming a key mode of human-computer interaction \cite{luger_like_2016}.
Prior research investigated the use of conversational agents in a variety of contexts (e.g., collaborative games \cite{ashktorab_effects_2021, ashktorab_human-ai_2020}, customer services \cite{ashktorab_resilient_2019}, journaling and reflection \cite{kocielnik_designing_2018}, productivity applications \cite{grover_design_2020}, \rv{and business documents \cite{jahanbakhsh_understanding_2022}}). 
However, the use of a conversational assistant for UX analysis has been unexplored. 
In this study, we seek to understand how best to design a conversational assistant that helps UX evaluators by investigating what questions evaluators had while conducting analysis and how they wanted to ask these questions.
\rv{This follows the methodology of prior work that characterized the
information needs and queries of a conversation assistant for business documents \cite{jahanbakhsh_understanding_2022}.}
Through the reciprocal dialogue between humans and AI, deeper cooperation may be established
\cite{battistoni_change_2021, jentzsch_conversational_2019}.
We foresee a collaboration mode in which the AI processes multimodal information from recordings to identify acoustic and textual cues from users \cite{fan_automatic_2020}, while the UX evaluator can synthesize the AI suggestions together with the context of the user's task to make informed judgments on usability problems. Thus, the different capabilities of AI and UX evaluators can complement each other to achieve robust results.

Prior studies found that when locating usability problems, UX evaluators considered what users are doing and saying, as well as how they say it (e.g., pauses, tone), while paying close attention to feelings, comments, and design recommendations from users \cite{fan_practices_2020, norgaard_what_2006}. 
These studies provided examples of information that UX evaluators extracted from the recordings, but it remains unknown what types of questions they would find helpful to ask an AI assistant instead of determining on their own, and what types of information they would trust and feel comfortable receiving from AI assistants.
Thus, our work takes a first step at understanding the range of questions that UX evaluators are interested in asking, which would inform the design of future conversational AI assistants for UX analysis. 

\begin{figure*}[tb]
  \centering
  \includegraphics[width=\linewidth]{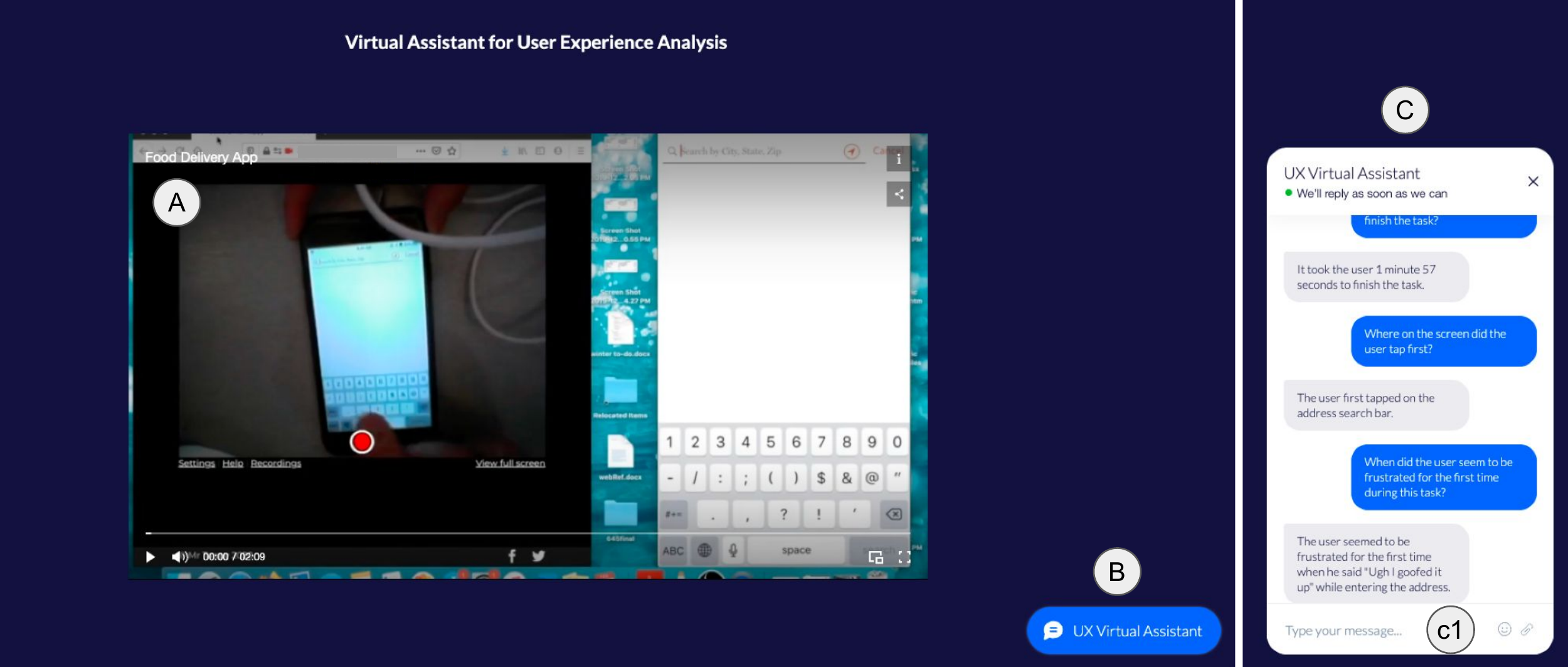}
  \caption{User interface of the text assistant: (A) Video player, (B) Chat bubble, (C) Chat thread that opens after the chat bubble is clicked, records the conversation between UX evaluators and AI assistant, and (c1) Chatbox to type questions.}
  \Description{Two screenshots showing the user interface of the text assistant: (A) Video player, (B) Chat bubble, (C) Chat window that opens after the chat bubble is clicked, records the conversation between UX evaluators and text assistant, and (c1) Chatbox to type questions.}
  \label{fig:text-assistant-UI}
\end{figure*}

\subsubsection{Text vs Voice Conversational Assistants}

Text chatbots have become the fastest growing communication channel \cite{moran_25_2022} and by 2022, 70\% of white-collar workers  will interact with conversational platforms on a daily basis \cite{goasduff_chatbots_2019}.
Similar to the increasing adoption of text assistants, voice assistants have also been on the rise \cite{kinsella_voice_2020}.
From 2018 to 2020, voice assistant usage on smartphones rose from 51.5\% to 56.4\%, while smart speaker ownership rose from 22.9\% to 34.7\% among U.S. adults \cite{kinsella_voice_2020}.
Respondents in a large-scale survey indicated that major reasons why they use voice assistants include hands-free interaction (55\%), it's fun (23\%), and speaking to the assistant feels more natural than typing (22\%) \cite{olmstead_nearly_2017}.
Prior studies strongly demonstrated differences in user behavior when participants used speech or text to interact with conversational interfaces \cite{kocielnik_designing_2018, luria_comparing_2017, kang_understanding_2017}. 
Thus, we sought to understand the benefits and drawbacks of these two interaction modalities in the context of UX evaluation. 

Researchers who developed a conversational agent for journaling and reflection found that text interactions were considered more familiar and efficient, whereas voice interactions have the potential to feel more interactive and engaging \cite{kocielnik_designing_2018}.
In the context of collaborative writing where writers received either written or spoken comments from reviewers, spoken comments were preferred and led to more positive perceptions of the reviewer \cite{neuwirth_distributed_1994}.
Other work compared queries to a movie recommendation system using voice versus typing, which showed that speaking led to longer queries that were more likely to contain subjective features than typing \cite{kang_understanding_2017}.
Furthermore, a higher proportion of spoken questions were labeled as ``conversational'' (i.e., as though the user was conversing with a human) \cite{kang_understanding_2017} and longer messages tended to be positively associated with engagement \cite{chae_effects_2020, van_heerden_potential_2017}, which suggest that voice interactions may be preferred. 
On the other hand, speaking resulted in longer time taken to ask questions, which reduced the efficiency of participants \cite{kang_understanding_2017}. 
Overall, prior work demonstrated trade-offs between efficiency and engagement when deciding between text or voice assistants. 
To balance these trade-offs, we must also consider the specific context of their usage.
While conversational assistants for workplace journaling and reflection may require them to feel more personal \cite{kocielnik_designing_2018}, our focus is on how conversational assistants can support UX evaluators with usability analysis.
In such settings, avoiding the disruption of work and improving efficiency may be more important since UX evaluators are often under time pressure \cite{kuang_merging_2022}. 
Thus, we seek to understand the differences in user behavior and user preferences between text and voice assistants in the specific task of reviewing usability test recordings.

\section{Design of the Conversational AI Assistant to Support UX Analysis}

In this section, we describe the design of the AI assistant in terms of its user interface and implementation as well as the Wizard of Oz approach.


\subsection{User Interface and Implementation}
In line with prior visual analytics tools to assist UX evaluators \cite{fan_vista_2020}, we followed the principle of \textbf{being simple and informative}. 
This approach ensured that UX evaluators can focus on interacting with the AI assistant without distractions from other interface elements since the UI only contains a video player and chat window (Fig. \ref{fig:text-assistant-UI} and Fig. \ref{fig:voice-assistant-UI}).
The UI is implemented as a web application, with the frontend built on top of React and Socket.IO and the backend built using Node.js and Socket.IO.

\subsubsection{Text Assistant}
The user interface for the text assistant contains two main interface components: (1) a video player for UX evaluators to review recordings (Fig. \ref{fig:text-assistant-UI}-A), and (2) a chat window that displays the conversation between UX evaluators and the text assistant (Fig. \ref{fig:text-assistant-UI}-C).
The video player contains typical playback controls including play/pause, volume up/down, a progress bar, and a full screen button. 
The chat window is collapsed by default into a chat bubble (Fig. \ref{fig:text-assistant-UI}-B) so that UX evaluators can focus on reviewing the video first, as this is a common playback strategy for analyzing recordings \cite{fan_vista_2020}. 
When they are ready to interact with the AI assistant, they can click to open the chat window and directly type into the chatbox at the bottom (Fig.  \ref{fig:text-assistant-UI}-c1).
All responses from the AI assistant are also displayed in the chat window. 

\subsubsection{Voice Assistant}

\begin{figure*}[tb]
  \centering
  \includegraphics[width=\linewidth]{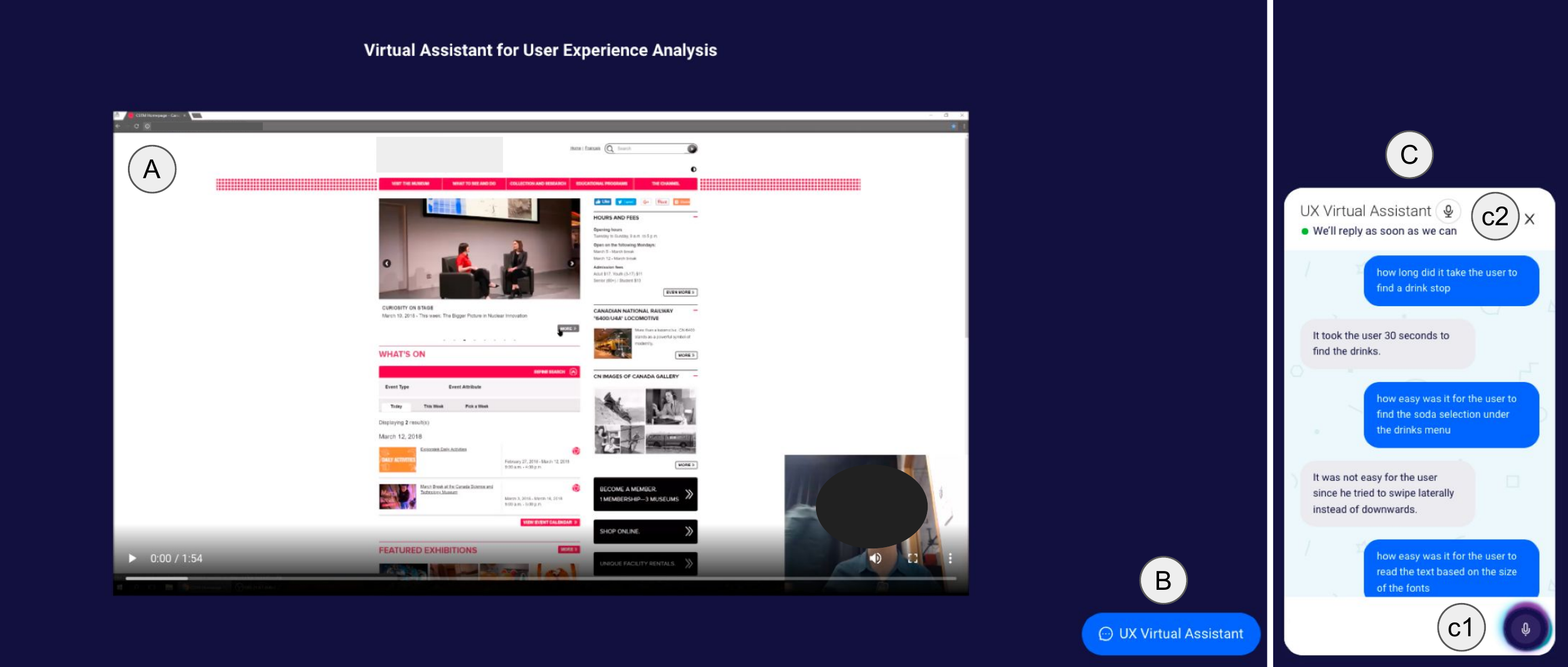}
  \caption{User interface of the voice assistant: (A) Video player, (B) Chat bubble, (C) Chat window that opens after the chat bubble is clicked, records the conversation between UX evaluators and voice assistant, (c1) Microphone icon that UX evaluators can click to enable speech-to-text transcription, and (c2) Circular icon that UX evaluators can click to mute the responses from the voice assistant.}
  \Description{Two screenshots showing the user interface of the voice assistant: (A) Video player, (B) Chat bubble, (C) Chat window that opens after the chat bubble is clicked, records the conversation between UX evaluators and voice assistant, (c1) Microphone icon that UX evaluators can click to enable speech-to-text, and (c2) Circular icon that UX evaluators can click to mute the responses from the voice assistant.}
  \label{fig:voice-assistant-UI}
\end{figure*}

Similar to the text assistant, the user interface for the voice assistant contains two main interface components: (1) a video player (Fig. \ref{fig:voice-assistant-UI}-A), and (2) a chat window (Fig. \ref{fig:voice-assistant-UI}-C).

This interface contains the same video controls and collapsed chat bubble by default. 
The differences are in the interaction with the AI assistant: 
in the text condition, UX evaluators can type into the chatbox to initiate conversation, whereas for the voice condition they can say \textit{``Hey UX assistant''} or click and hold the microphone icon (Fig. \ref{fig:voice-assistant-UI}-c1) to record their question.
Once the evaluator's microphone is on, the system automatically pauses the video. 
The blue and purple halos around the microphone icon also begin pulsing, which acts as visual feedback that the mic is on and indicates that they can start speaking.
The responses from the AI assistant are shown both as a written message in the chat window as well as spoken aloud.
A female-sounding voice was used in the text-to-speech algorithm, which is consistent with the most common commercially available voice assistants \cite{cambre_one_2019}. 
When the responses start playing, the video is automatically paused.
However, if UX evaluators prefer not to hear the responses out loud, they can click the circular icon on the top of the chat window (Fig. \ref{fig:voice-assistant-UI}-c2:\icon{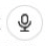}) to mute the voice assistant. 
The icon then changes to\icon{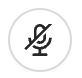} to indicate that the assistant has been muted. 
We chose to display the responses both in voice and text so that UX evaluators have the choice to mute responses if it was distracting.
Furthermore, prior work showed that 70\% of literate and semi-literate users preferred having both audio and text responses from voice assistants instead of audio-only \cite{jain_farmchat_2018} and providing relevant visual feedback in addition to voice responses resulted in higher perceived usefulness \cite{baeza_perceived_2019}.

In order to transcribe UX evaluators' verbalizations into text, the system used the React Speech Recognition package \cite{react_speech_recognition_2022}, which relied on the Web Speech API under the hood. 
The Web Speech API uses the speech recognition available on the device and is supported on Google Chrome browsers \cite{web_speech_api_2022}.
The main challenge for the implementation of the voice assistant was to enable the microphone permissions for voice input since it required an SSL certificate on the system's domain.
As soon as a UX evaluator connected to the web application, a pop-up notification asked for their microphone permissions.
Once granted, they were able to interact with \rv{the voice-to-text engine.}

\subsection{Wizard of Oz Design}
As it is still challenging to leverage state-of-the-art AI algorithms to accurately detect usability problems and provide natural language responses \cite{fan_human-ai_2022}, we adopted a Wizard of Oz design to simulate conversational agents so that we could better focus on answering our research questions. Wizard of Oz has been commonly used to circumvent technical limitations in prior research (e.g., \cite{medhi_thies_how_2017, shamekhi_face_2018, luria_comparing_2017}). 

\subsubsection{Acting as the AI Assistant}

In the design probes, one moderator acted as the AI assistant by responding to the questions from participants.
For the text condition, the moderator received questions from participants \rv{in a chatbox that} was only accessible via an administrator account and password. 
The moderator then typed a response, which was displayed in the chat windows on the participants' browsers (Fig. \ref{fig:text-assistant-UI}-A and Fig. \ref{fig:voice-assistant-UI}-A).

In the voice condition, the moderator could hear the participant through videoconferencing software that was external to the UI.
To ensure the consistent activation of the voice assistant, once the moderator heard the activation command of \textit{``Hey UX assistant''}, they remotely turned on the participant's microphone.
The moderator also received the transcribed questions from the participant in \rv{the chatbox} and typed responses. 
To account for occasional errors in the speech-to-text algorithm, the moderator responded to the voiced question as they understood it.
For example, transcriptions like \textit{``what is the time taken to find a dress''} were assumed to be \textit{``what is the time taken to find the address.''}

\subsubsection{Capabilities of the AI Assistant}
\label{sec:capabilities}
To determine the AI assistant's capabilities, we referred to existing literature and commercial analysis platforms.
Based on prior work that utilized machine learning to automatically extract acoustic (e.g., pitch, loudness, and speech rate), textual (e.g., negations, questions, sentiments), and visual (e.g., scrolling speed, scene breaks) features directly from the recordings \cite{fan_automatic_2020, fan_vista_2020, soure_coux_2021}, we determined that the AI assistant should be able to answer questions relating to these features.
Since existing commercial usability testing tools have the capability to count the number of clicks through heatmaps and time spent on a certain page \cite{usertesting_usertesting_2021, userzoom_userzoom_2021}, we manually extracted this information from the recordings used in the design probe.
Furthermore, we assumed that the AI assistant would have knowledge of higher-level semantic features, such as user actions or emotions that can be detected via machine learning recognition techniques \cite{fan_human-ai_2022}.
In addition to extracting the \rv{information above}, two researchers also independently analyzed the recordings to identify usability problems and derive redesign recommendations before collaborating to consolidate the list, which follows best practices for usability analysis \cite{hertzum_evaluator_1998, kuang_merging_2022}.
\rv{Thus, the AI assistant could provide responses such as \textit{``Yes, I believe there is a usability problem because the user said oops, I clicked on the wrong button.''}}
If participants asked questions that fell outside of the capabilities of the AI assistant, they received a standard response of \textit{``Sorry, I don't know the answer to this question''}, as prior work recommended that conversational agents should express the gaps in their knowledge \cite{macskassy_conversational_1996}.
\section{User Study}

We conducted a between-subjects Wizard-of-Oz design probe \rv{with 10 participants using the text assistant and 10 participants using the voice assistant} to collect a dataset of questions that UX evaluators would ask during usability analysis. 

\subsection{Participants and Apparatus}
We recruited 20 participants (14 females, and 6 males) through social media and mailing lists. 
They were UX researchers ($N=14$), senior UX researchers ($N=2$), UX research interns ($N=2$), and UX/HCI graduate students ($N=2$). 
Participants self-reported having 1-13 ($M=3.6, SD=2.5$) years of prior UX experience.
They were randomly assigned to either the text assistant or the voice assistant, with 10 participants each. The average years of UX experience for text and voice conditions were 3.7 and 3.5 years ($SD=3.4, 1.4$) respectively. 
The median of their self-reported familiarity with usability analysis was the same for both conditions: ``4 - very familiar'' (on a scale of 1-5). 
\rv{Mann-Whitney U tests found} no significant differences in the years of UX experience or familiarity with usability analysis between conditions. 
All participants had prior experience interacting with text chatbots and voice assistants like Apple Siri, Google Assistant, and Amazon Alexa. 
All participants completed the study remotely using their own computers to access the web application while communicating with the moderator through video-conferencing software.

\subsection{Study Videos}
\rv{Since there is currently no established taxonomy of products or tests that need to be covered during studies of UX analysis tools, we selected some examples to prompt analysis, which follows prior work (e.g., \cite{fan_vista_2020, fan_human-ai_2022, soure_coux_2021}).}
We used two recorded usability test recordings collected from prior research projects.
\rv{Although two videos can not be representative of all usability tests or tasks, we covered two common digital interfaces (website and smartphone app) and two user groups (female young adult and male older adult).}
In the website video (length: 1 minute 54 seconds), a young adult was asked to search for an event appropriate for an 11-year-old on a science and technology museum's website.
In the app video (length: 2 minutes 9 seconds), an older adult was asked to find a grocery store by entering an address and adding 10 bottles of Coke to the cart on a food delivery app.
These videos were also selected since they each contained at least three usability problems, which provided opportunities for participants to conduct analysis and ask questions to the AI assistant. 
\rv{During the study, all participants watched both videos, but only used one modality to interact with the AI assistant.}

\subsection{Procedure}
Participants connected to the moderator via videoconferencing software. 
They were given a short tutorial about either the text or the voice assistant, including how to interact with them (e.g., type their question in the chatbox or say \textit{``Hey AI assistant''}). 
The following prompt was given to participants to mimic the goal-oriented nature of the task: 

\begin{displayquote}
\textit{``In this scenario, you are a UX evaluator who is tasked with analyzing 2 usability test recordings. You are expected to report the identified usability issues to your team afterward. While reviewing the recordings, please [type/say] any questions that you believe would be helpful to your analysis to the AI assistant. \rv{The AI assistant is limited to answering questions about the specific video you are currently analyzing, including observable actions and verbalizations of the users. It has basic UX functionalities like identifying usability problems and heuristics-based recommendations, so it may respond with ``I don't know'' to some of your questions. However, you are encouraged to ask any questions relevant to your analysis regardless of whether an answer is provided or not.}''}
\end{displayquote}

Participants were able to ask any questions about the study tasks and the web application before proceeding. 
Then the moderator explained the scenario in the first recording (the order of website and app videos was counterbalanced between participants) and asked the participants to \rv{proceed with analysis}.
After participants finished analyzing the first video, the moderator explained the scenario in the second recording.
Once participants finished analyzing both videos, they completed a short Likert scale survey and semi-structured interview about their experience. 
Participants spent 10-15 minutes analyzing each short video, which resembles the time constraints that UX evaluators have in industry \cite{fan_practices_2020, kuang_merging_2022}. 
All sessions were video-recorded and lasted 35-45 minutes. Participants were compensated for their time.
\rv{In order to respond to participants as quickly as possible, the moderator had access to the transcript and a list of usability problems for the videos, as well as other features with the associated timestamps that were generated by prior analysis (Sec \ref{sec:capabilities}).}
The moderator remained the same for all participants and all videos so that the content and semantics of the responses from the simulated AI assistant would be consistent. A post hoc check was done by examining the conversation records, especially responses to common questions, and confirmed that they were uniform across participants.

\section{Results}
In this section, we present the findings on the categories and order of questions ask (RQ1), differences in questions between text and voice assistants (RQ2), and participant perceptions of text and voice assistants (RQ3). Participant $x$ in the text condition is labeled P$x-T$ and participant $x$ in the voice condition is labeled P$x-V$.

\begin{table*}[tb]
  \centering
  \caption{Categories of questions asked by participants.}
  \label{tab:typesQuestions}
  \small
  \begin{tabular}{p{2cm} p{3cm} p{2cm} p{8cm}}
    \toprule
    \textbf{Category} & \textbf{Subcategory} & \textbf{Number of Questions (\%)} & \textbf{Example Question} \\
    \midrule
    \rv{\multirow{3}{2cm}{User actions}} & Current user's actions & 94 (28.9\%) & ''How many clicks did the user have to go through to reach the target page?'' \\
    & Other users' actions & 4 (1.2\%) & ``Did other participants also open the wrong link?'' \\
    \midrule
    \rv{\multirow{5}{2cm}{User mental model}} & User perceptions & 30 (9.2\%) & ``Did the participants find it easy to navigate the website?'' \\
    & User emotions & 24 (7.4\%) & ``When did the participant seem to be confused for the first time during this task?'' \\
    & Reasons for user actions & 16 (4.9\%) & ``Why did the user always open a new tab instead of staying on the same page?'' \\
    \midrule
    \multirow{5}{2cm}{Help from AI assistant} & Assistant suggestions & 65 (20.0\%) & ``Can you tell me what recommendation we can provide for the usability issue in the beginning?'' \\
    & Assistant capabilities & 11 (3.4\%) & ``Can you take notes for later or add timestamps to the video?'' \\
    & Search engine & 8 (2.5\%) & ``What's the principle related to help and documentation?'' \\
    & Volume control & 1 (0.3\%) & ``Can you make the participant's voice louder?'' \\
    \midrule
    \multirow{3}{2cm}{Product and task information} & Product information & 35 (10.8\%) & ``Are there filters for the user to search for appropriate events at the museum?'' \\
    & Task information & 19 (5.8\%) & ``What is the ideal path to complete this task?'' \\
    \midrule
    \multirow{2}{2cm}{User demographics} & User background & 16 (4.9\%) & ``Does this user have prior experience with food delivery apps?'' \\
    & Other users' background & 2 (0.6\%) & ``What is the inclusion criteria for participants?'' \\
    \bottomrule
\end{tabular}
\end{table*}

\subsection{Categories and Order of Questions Asked (RQ1)}

In total, our study collected a \textbf{dataset of 325 questions}, of which 181 questions (56\%) were asked to the text assistant and 144 questions (44\%) to the voice assistant.
\rv{Out of 1511 total words in the voice condition, 32 (2.1\%) were inaccurately transcribed by the React speech recognition package. 
Once participants noticed a transcription error, they repeated the same question but spoke more slowly and clearly. Thus,} fourteen questions from the voice condition were removed since they were repetitive. 
Each question was coded by the researchers independently and then consolidated in a group discussion. We came up with 12 labels, which were then grouped into five larger categories. 

\subsubsection{Categories of Questions}

\begin{figure*}[htbp]
    \centering
    \includegraphics[width=0.45\linewidth]{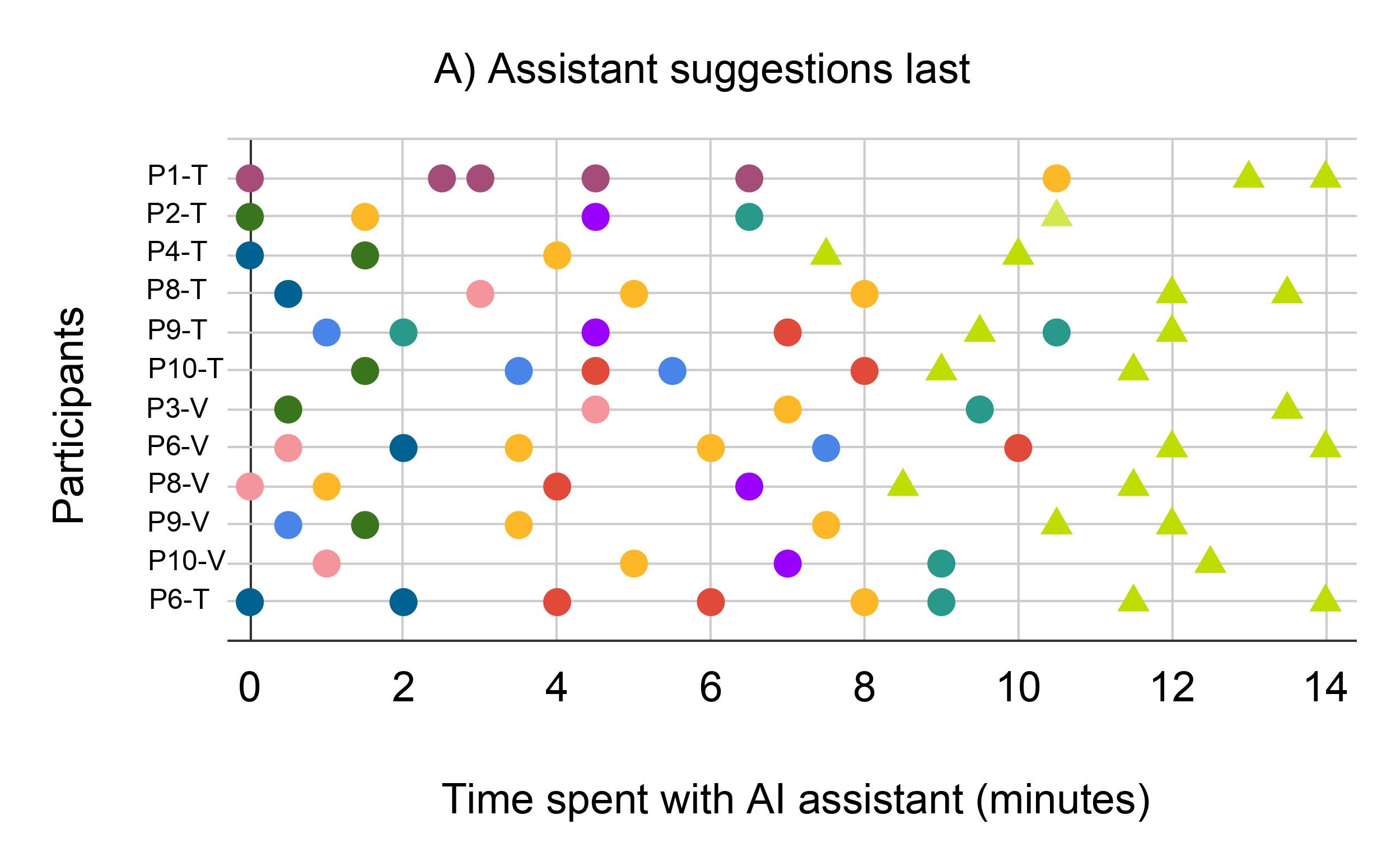}
    \enspace
    \includegraphics[width=0.45\linewidth]{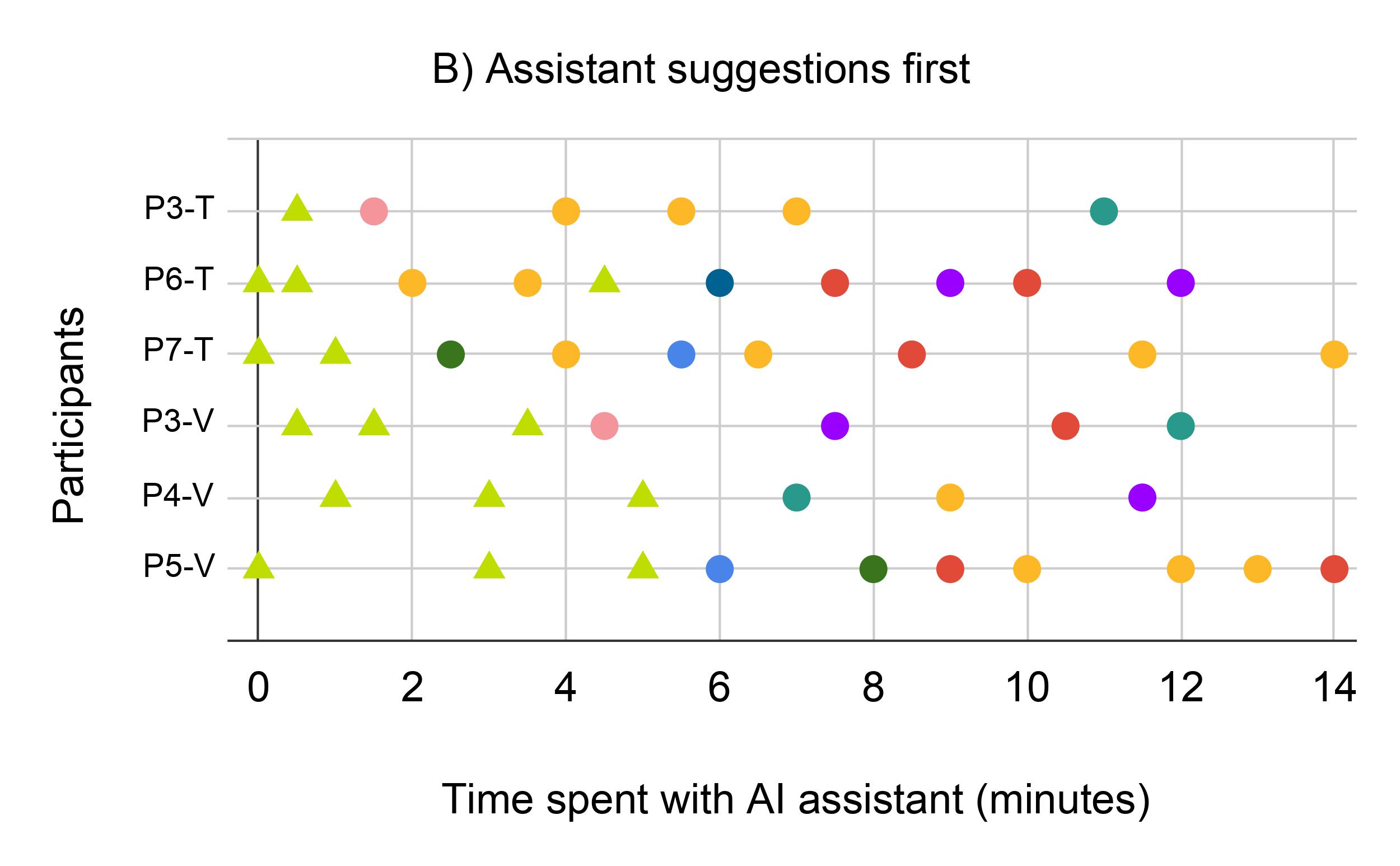}
    \enspace
    \includegraphics[width=0.8\linewidth]{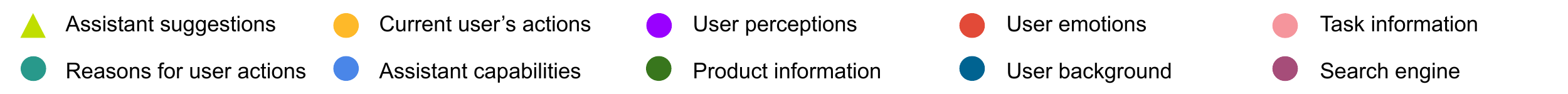}
    \caption{\rv{Timeline plots} showing the order of the question categories as participants spent time interacting with the AI assistant: (A) Assistant suggestions last, (B) Assistant suggestions first. \rv{(Each line represents one participant.)}}
    \Description{Two timeline plots showing the order of the question categories as participants spent time interacting with the AI assistant: First plot shows assistant suggestions were asked at the end, while the second plot shows assistant suggestions asked at the beginning.}
\label{fig:question-order}
\end{figure*}

Table \ref{tab:typesQuestions} \rv{lists} the categories, subcategories, and the corresponding number of questions that participants asked the AI assistant. 
Our results show that the expected functionalities of the AI assistant fell in the following categories: 
(1) user actions, \rv{(2)} user mental model, \rv{(3)} help from AI assistant, \rv{(4)} product and task information, and \rv{(5)} user demographics. 

\rv{
\textbf{User actions:} Almost one-third (98 or 30.2\%) of all questions belonged to this category.}
Their questions included \textbf{current user's actions} made on the interface (e.g., the number of clicks the user made to reach the target page or duration of time spent on a particular page).
Some participants were also interested in \textbf{other users' actions} to determine the frequency of a problem (e.g., \textit{``Did other participants also open the wrong link?''}). 

\rv{\textbf{User mental model:} This category consisted of 70 (21.5\%) questions with the subcategories of \textbf{user perceptions}, \textbf{user emotions}, and \textbf{reasons for user actions}. 
Many participants asked if the user seemed confused or found it easy to navigate the interface.}
The AI assistant answered these questions based on the speech and facial expressions of the users in the usability recordings.
\rv{For example,} when P8-T asked \textit{``When did the user seem to be frustrated for the first time during this task?''}, the AI assistant responded with \textit{``The user first seemed frustrated when he said ``Ugh I goofed it up'' while entering the address.''}
\rv{Participants also asked for reasons behind certain questions (e.g., \textit{``Why did the user always open a new tab?''} -P3-T). In these cases, the AI assistant did not provide speculative answers if the user did not explicitly verbalize their reasons.}

\textbf{Help from AI assistant:} This category contained 85 questions (26.2\%) and includes the subcategories: \textbf{assistant suggestions}, \textbf{assistant capabilities}, \textbf{search engine}, and \textbf{volume control}. 
For the assistant suggestions subcategory, participants asked the AI assistant for its opinions on whether a usability problem existed or if a specific interface changed, as well as for design recommendations on how to address usability issues. 
For example, P9-T asked \textit{``For the options at the top, do you think we should add a drop-down menu that shows the sub-options?''}. 
\rv{In the post-task interviews,} some participants mentioned that they were testing the limitations of the AI assistant by asking harder questions or considered it a coworker who could provide suggestions on interface changes to address usability problems. 
In addition, participants asked the AI assistant if it had specific capabilities such as note-taking.
They also used the AI assistant as a search engine like Google to look up definitions of Nielsen's heuristics \cite{nielsen_10_1994} and asked the AI assistant to change the volume of the video.
In these cases, the AI assistant answered that it could not help since it was limited to natural language responses, but this shows that some participants had a desire for more comprehensive assistance with other tasks during analysis.

\textbf{Product and task information:} This category with 53 questions (16.6\%) included \textbf{background information} on the website or app as well as \textbf{information about the specific task}.
\rv{Questions in this category show that in addition to knowledge of the video content, participants would like the AI assistant to have background knowledge such as whether filters were available on the website or what the ideal path to complete the task was since these are factors that they consider when determining the occurrence of a usability problem.}

\textbf{User demographics:} The last category with 18 questions (5.5\%) contained questions about this \textbf{particular user} and \textbf{other users} in the study, like the inclusion criteria for the study. 
It is important to note that participants often required additional background information to better understand how the user's personal, social, or technological experience may have influenced the task and see how frequently it occurred across multiple users.
For example, P2-T asked \textit{``how many other participants opened new tabs during the study?''} to understand the scale of the issue.

\subsubsection{Order of Category Occurrence}
To identify common sequences for the order, we plotted the dataset of questions with different colors representing each subcategory. 
We found that there were two main strategies for interacting with the AI assistant: (1) assistant suggestions last \rv{(Fig. \ref{fig:question-order}-A)}, and (2) assistant suggestions first \rv{(Fig. \ref{fig:question-order}-B)}.
Since the videos were about 2 minutes long, all participants finished watching the whole video before engaging with the AI assistant. 


The first strategy was to ask for \textbf{assistant suggestions last}, which was adopted by twelve participants. 
This strategy consisted of gathering context on the user before asking for suggestions from the AI assistant.
\rv{In Fig. \ref{fig:question-order}-A, the lime triangles are clustered on the right side, which indicates that these questions were asked later in the session.}
We also found that certain categories like \textbf{product information} and \textbf{user background} were more likely to occur at the beginning of the analysis session.

In contrast to the first strategy, the second strategy was to ask for \textbf{assistant suggestions first} ($N=6$). 
\rv{In Fig. \ref{fig:question-order}-A, the lime triangles are clustered on the left side, which indicates that these questions were asked later in the session.}
Post-task interviews revealed that participants wanted to first understand the capabilities of the AI assistant to decide whether it would be useful for them before asking for more detailed information to complete their analysis.
\rv{For the participants who did not follow either strategy, there were no obvious patterns in their sequence of questioning.}

\subsection{Differences in Questions between Text and Voice Assistants (RQ2)}

Since prior work showed that participants interacted with text and voice assistants differently in various tasks (e.g., longer queries to voice assistants \cite{kang_understanding_2017}), we were interested in understanding whether those differences remained consistent for UX evaluation.

\begin{figure}[h]
  \centering
  \includegraphics[width=\linewidth]{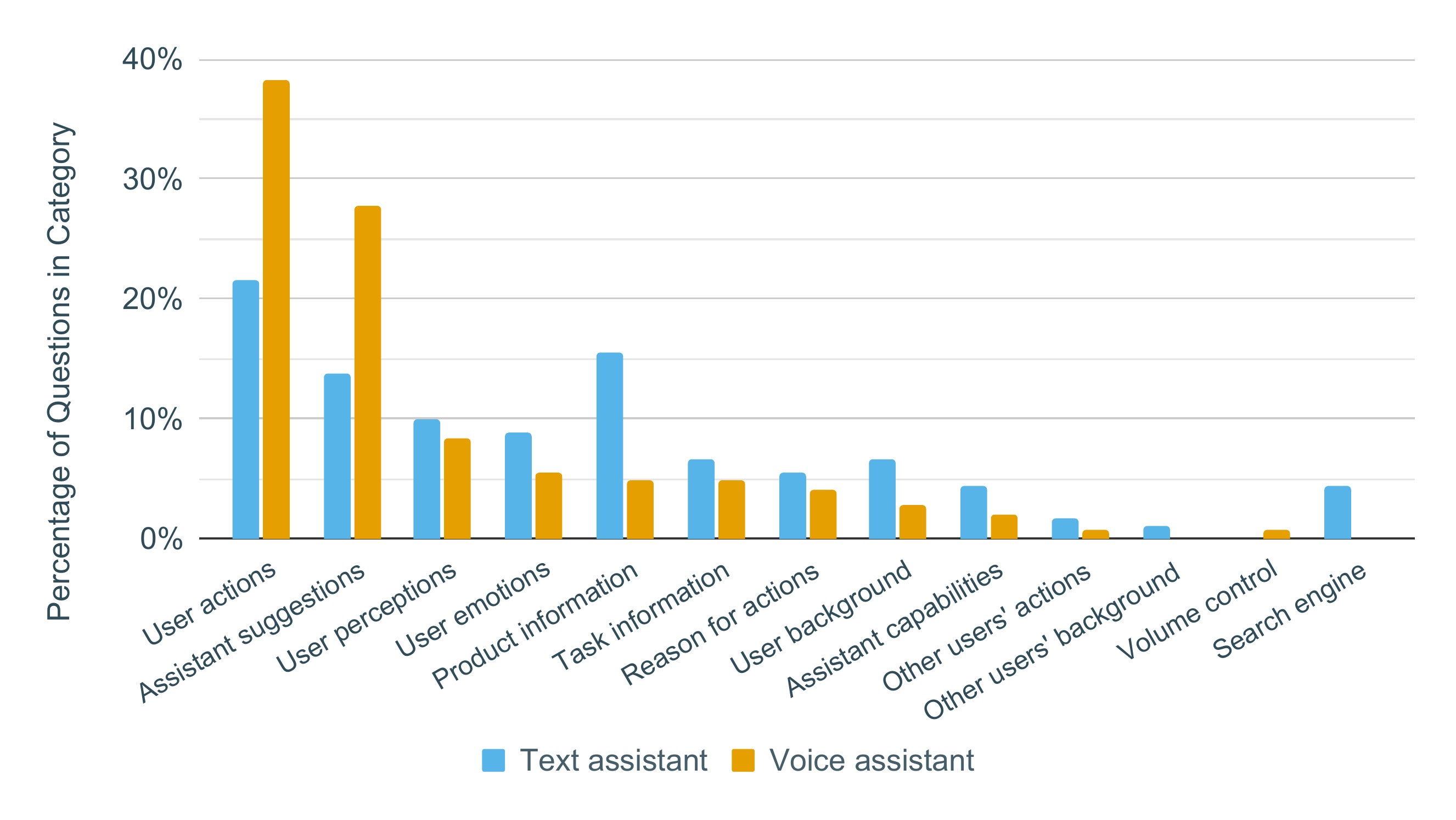}
  \caption{Bar chart showing the percentage of questions in each category separated by text and voice assistants.}
  \Description{Bar chart showing the percentage of questions in each category. The top three for text assistants are user actions (22\%), product information (15\%), and assistant suggestions (14\%) while the top three for voice assistants are user actions (38\%), assistant suggestions (28\%), and user perceptions (8\%).}
  \label{fig:question-categories}
\end{figure}

\subsubsection{Categories of Questions}

Fig. \ref{fig:question-categories} shows the percentage of questions in each category separated by text and voice assistants.
The top three categories for text assistants are \textbf{user actions} (22\%), \textbf{product information} (15\%), and \textbf{assistant suggestions} (14\%) while the top three categories for voice assistants are \textbf{user actions} (38\%), \textbf{assistant suggestions} (28\%), and \textbf{user perceptions (8\%)}.
The division of categories shows that the categories for the text assistant were more evenly distributed than the voice assistant.
We also observed two interesting patterns: (1) user actions and assistant suggestions are two common categories across both conditions, and (2) product information was asked more often to the text assistant than the voice assistant, which we discuss in Sec \ref{sec:categories}.

The \textbf{search engine} category was unique to the text assistant and contributed by one participant, \rv{P1-T, who used the assistant to look up definitions of UX terms.}
The \textbf{volume control} category only occurred for the voice assistant and was also based on one particular participant. 
P2-V asked the assistant to make the volume louder because he
\textit{``assumed that it could help with changing the sound''} since he was already speaking to the assistant.

\subsubsection{Number of Questions}

\begin{table}[tb]
  \centering
  \caption{Number of questions asked per participant}
  \label{tab:numberQuestions}
  \small
  \begin{tabular}{ll}
    \toprule
    \textbf{Number of Questions} & \textbf{Mean (SD)} \\
    \midrule
    Text assistant - Website & 10.4 (3.6) \\
    Text assistant - App & 7.7 (1.6) \\
    \midrule
    Voice assistant - Website & 7.7 (3.1) \\ 
    Voice assistant - App & 7.2 (1.8) \\
    \bottomrule
\end{tabular}
\end{table}

Table \ref{tab:numberQuestions} shows the average number of questions asked by participants during the 10-15 minutes that they spent analyzing each video.
Across all four videos, participants asked on average 8 questions ($SD=3$).
\rv{Since participants asked numerous questions in a short time and at a regular pace (as shown in Fig~\ref{fig:question-order}), they seemed to be actively engaged with the AI assistant throughout their analysis.}
\rv{We used Shapiro-Wilk to check the normality of the collected data, then conducted a two-way ANOVA with the factors being modality (voice or text) and interface (website or app).
The effect of modality was statistically significant ($F_{1,36} = 4.9, p < .05, \eta_{p}^{2} = 0.1$), while the effect of interface was not statistically significant ($F_{1,36} = 2.6, p > .05, \eta_{p}^{2} = 0.07$). There was also no interaction effect between modality or interface. 
Post-hoc comparisons with Bonferroni correction revealed that participants asked significantly more questions using the text assistant when analyzing the website video than both videos using the voice assistant. 
}
This finding may be related to the text assistant having higher perceived efficiency, which is described in Sec \ref{sec:efficiency}.

\subsubsection{Length of Questions}

\begin{figure}[h]
  \centering
  \includegraphics[width=\linewidth]{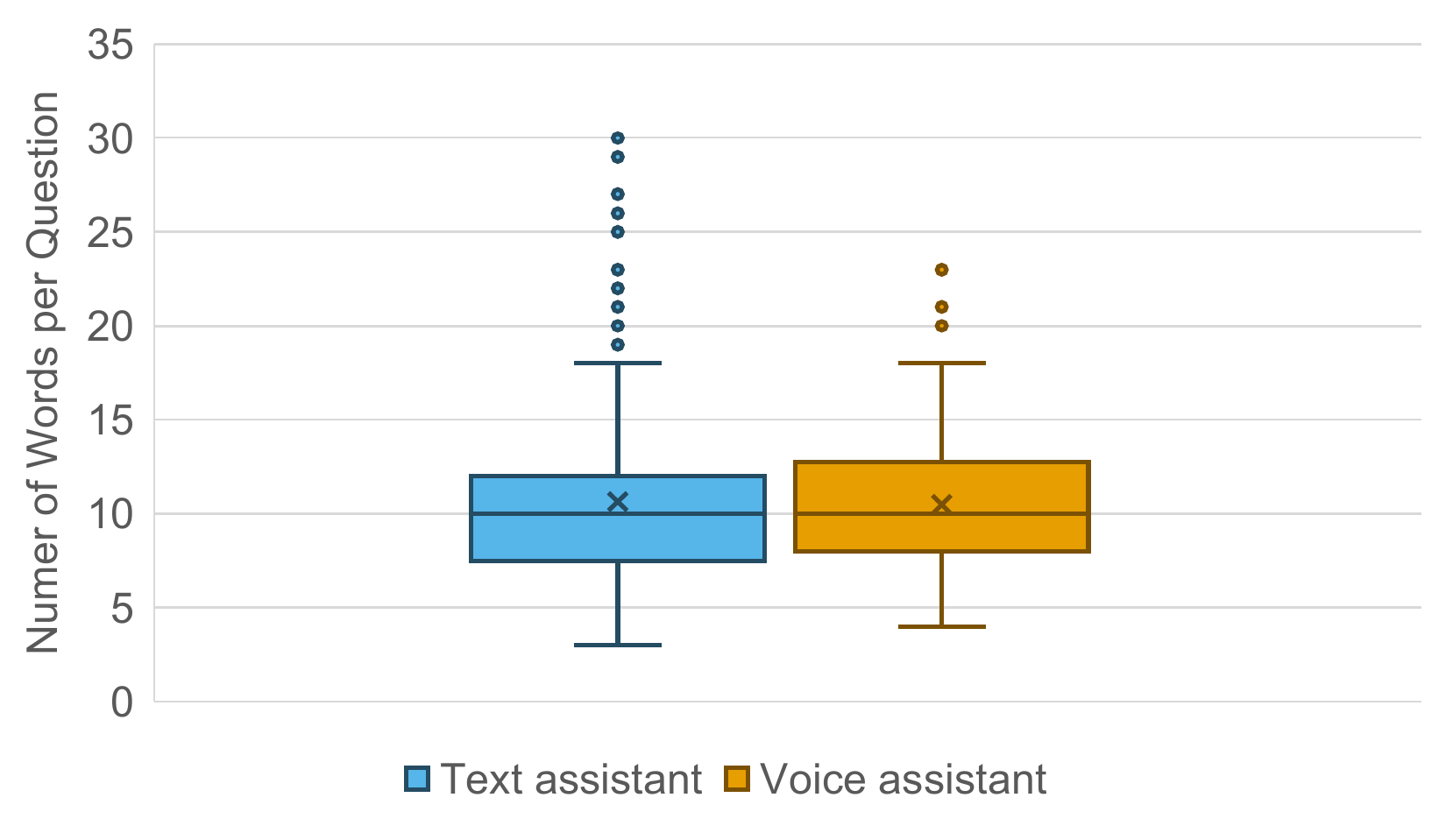}
  \caption{Box plot showing the distribution of the number of words per question for text and voice assistants.}
  \Description{Box plot showing the range and inter quartiles of the number of words in each question. The box plot for text assistant has a lower limit of 3 and an upper limit of 18 with a median of 10. The box plot for voice assistant has a lower limit of 4 and an upper limit of 20 with a median of 10.}
  \label{fig:number-words}
\end{figure}

Fig. \ref{fig:number-words} shows the distribution of the number of words per question for text and voice assistants. 
For the voice condition, the three words in the activation command \textit{``Hey UX Assistant''} were not counted as part of the question. 
The range for the text assistant is 3 to 30 words, while the range for the voice assistant is 4 to 23 words. The average words per question were 11 ($SD = 5$) and 10 ($SD=4$) respectively. 
There were no significant differences between conditions. 
However, the box plots show that there were more outliers---dots that lay above the upper limit of the whisker which is calculated as $Q3 + 1.5*IQR$---in the text condition (5.5\% of all typed questions) than the voice condition (2.1\% of all voiced questions).
When examining the long questions \rv{for both conditions}, we found that participants provided an \textbf{observation from the video} prior to asking the question.
For example, P4-T typed \textit{``It took for the user 1 minute 40 seconds to reach the destination. What's the ideal time expected for a user to reach the target page in this scenario?''} which contained 29 words.
Other participants provided their \textbf{rationale for a suggestion}, such as P9-T who wrote \textit{``Should the food delivery app track the history of what users search? If so, that might be helpful in the future to find the restaurant again''}, which contained 26 words.

\subsection{Participant Perceptions of Text and Voice Assistants (RQ3)}

\begin{figure}[h]
  \centering
  \includegraphics[width=\linewidth]{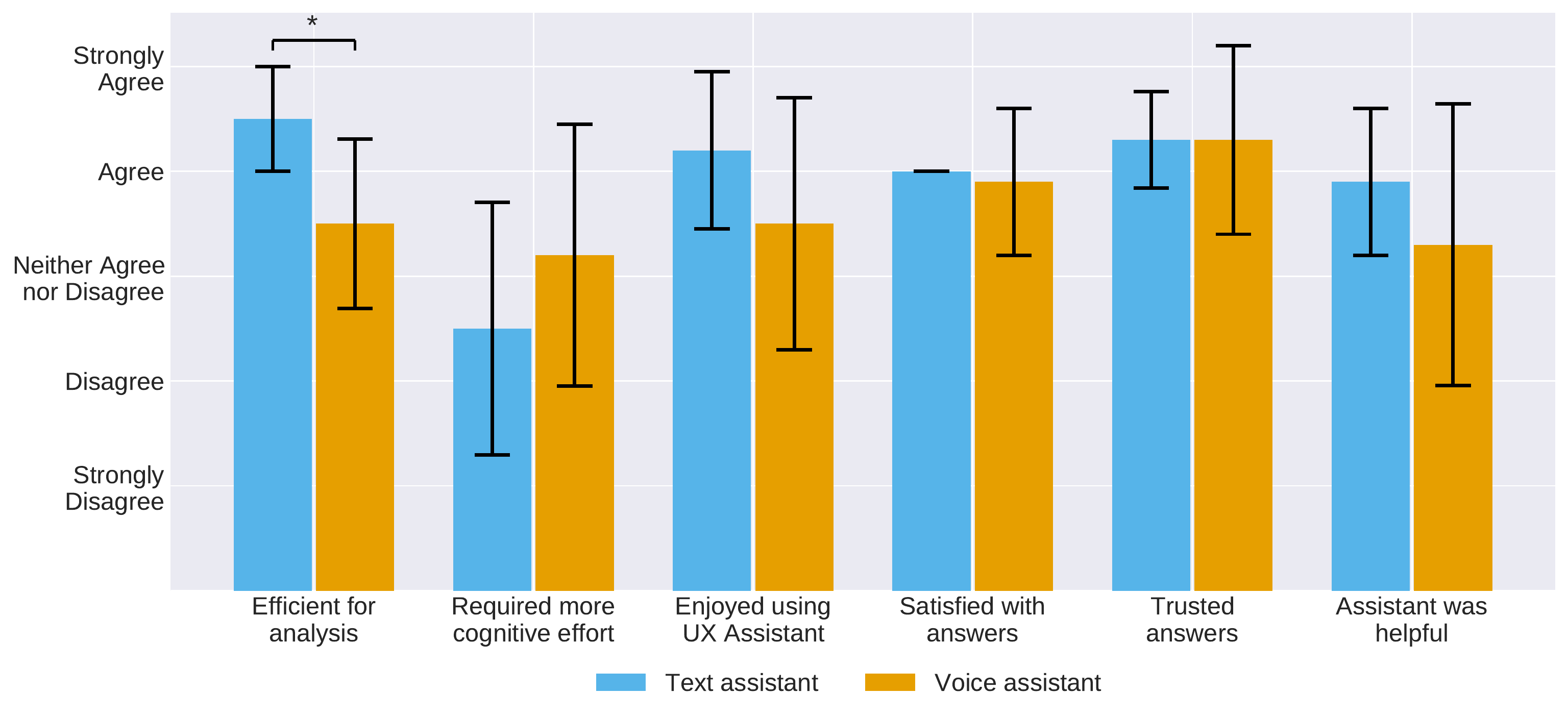}
  \caption{Bar chart showing the survey responses for both the text and voice assistants. (* $p < .05$)}
  \Description{Bar chart showing the survey responses for both the text and voice assistants. Text assistant was rated as significantly more efficient to use than the voice assistant.}
  \label{fig:survey-responses}
\end{figure}

Fig. \ref{fig:survey-responses} shows the Likert scale ratings of the text and voice assistants.
\rv{We conducted the Mann-Whitney U test since the ratings were non-parametric independent samples, and calculated the effect size using the formula: $r = \frac{|z|}{\sqrt{n}}$ \cite{datatab_mann-whitney_2022}.}

\subsubsection{Efficiency of Use} \label{sec:efficiency}
There was a significant difference for one out of the six survey responses: the text assistant was rated as significantly more efficient to use during analysis than voice assistant ($p < .05$).
\rv{The effect size for perceived efficiency was $r=0.54$\footnote{\rv{$r>0.5$ is considered large for Mann-Whitney Test \cite{datatab_mann-whitney_2022}.}}, which may explain} why more questions were asked to the text assistant than the voice assistant. 
Participants strongly agreed or agreed that they felt it was efficient to use the text assistant ($Md=4.5, IQR=1$). 
P8-T mentioned that \textit{``delivering information in this format is most efficient.''}
On the other hand, fewer participants agreed that the voice assistant was efficient ($Md=3.5, IQR=1$). 
P8-V mentioned that \textit{``The voice response interfered with the recording itself and wasn't very efficient.''}
However, other participants appreciated the convenience of the voice assistant: \textit{``I liked the convenience of speech-to-text, I can type notes on the usability problems while saying my questions which helps me multitask''} -P7-V.
The ability to multitask was also brought up by P2-V, P3-V, and P9-V. 

\subsubsection{Cognitive Effort}
Although there were no significant differences, participants generally rated the voice assistant as requiring more cognitive effort ($Md=4, IQR=2$) than the text assistant ($Md=2, IQR=1$).
P4-V said that \textit{``I had to actively think about how to phrase the question in the simplest way before I asked it so that the assistant could understand me.''}
P9-V also mentioned that \textit{``I tried to speak more slowly and succinctly so that it could accurately transcribe my question.''}
The moderator in the study observed that almost half of the participants in the voice condition exhibited this behavior, but there were also two participants in the text condition who spent extra effort to find the right wording.
They edited their questions multiple times in the chatbox before sending it and because they tried to \textit{``translate questions into the simplest grammatical form so that the assistant could understand''} -P2-T.

\subsubsection{Enjoyment}
Overall, participants enjoyed using the text assistant ($Md=4, IQR=1$) and voice assistant ($Md=3.5, IQR=2.5$). 
P4-T said that \textit{``using this would improve my productivity, it's great to feel like there is someone else who can discuss with me.''}
Similarly, P9-V mentioned \textit{``I enjoyed talking to the assistant, it felt like I was actually having a discussion with a research assistant or a colleague, although some responses were a little slow.''} 
Similar feedback from other participants confirm our hypothesis that presenting the extracted information through a conversational interface feels natural and collaborative. 

\subsubsection{Satisfaction with Answers from the AI assistant}
Overall, participants agreed that they were satisfied with the answers provided by both the text assistant ($Md=4, IQR=0$) and voice assistant ($Md=4, IQR=0.75$). 
P5-T liked that \textit{``the assistant doesn't show any emotions, it's straightforward at providing the desired information on the video.''}
On the other hand, P8-V mentioned that \textit{``the answers were on the simpler side, I'm getting useful information from the assistant but it can elaborate a bit more.''}

\subsubsection{Trust in the Answers}
Participants agreed that they trusted the answers provided by both the text assistant ($Md=4, IQR=0.75$) and voice assistant ($Md=4.5, IQR=1$). 
Some participants seemed surprised when it was revealed that this study was a Wizard-of-Oz design. 
P3-V mentioned \textit{``I didn't even think about it, I just fully trusted the responses.''}
Others calibrated their trust based on the responses they received. 
P2-T said that \textit{``Actually my trust went up when the assistant responded that they don't know something, since it seemed to recognize its limitations.''}
This feedback shows that it is important to communicate to participants the limitations of the system, especially when they are unfamiliar with it. 
On the other hand, P6-T was more cautious about the responses: \textit{``I liked that it provides the quantitative metrics like clicks on the website but I'm not sure how it was calculated, the assistant should explain the algorithm it used to count the clicks.''}
Related feedback from other participants show that in addition to directly answering a question, participants felt it was important to know how that data was captured.

\subsubsection{Perceived Helpfulness}
Although there were no significant differences, participants generally rated the text assistant as being more helpful ($Md=4, IQR=0.75$) than the voice assistant ($Md=3, IQR=2.5$).
Participants especially appreciated the AI assistant's help with more factual and objective information, such as counting the number of clicks or calculating the time taken to complete a subtask, which were tedious aspects of their analysis.
In addition, the chat window was helpful since P4-T mentioned that \textit{``having the conversation thread visible is useful for me to refer back to later and retrace the logic of my analysis.''} 

\section{Discussion}

\subsection{Support from the AI assistant}

\subsubsection{Categories of Questions} \label{sec:categories}

Through our study, we identified \rv{five} main categories of questions that were asked to the AI assistant: \rv{\textbf{user actions} (30.2\%), \textbf{user mental model} (21.5\%)}, \textbf{help from AI assistant} (26.2\%), \textbf{product and task information} (16.6\%), and \textbf{user demographics} (5.5\%). 
The most frequent category was \textbf{user actions and mental model}, which made up 47.5\% of all questions in the text condition and 56.9\% in the voice condition.
The large proportion in this category makes sense as prior work found that the majority of UX evaluators felt it was helpful to know what users are doing (94\%), what users are saying (86\%), and how they are saying it (76\%) when identifying usability problems \cite{fan_practices_2020,fan_concurrent_2019}.
User actions were also the type of information most frequently extracted in prior work on visual analytics tools for UX analysis (e.g., ML-driven textual and acoustic features that showed what the user said and visual features that showed scrolling speed \cite{fan_vista_2020, soure_coux_2021, fan_human-ai_2022}).

The \textbf{help from AI assistant} category made up 22.7\% of all questions in the text condition and 30.6\% in the voice condition. 
It required more subjective responses and showed that participants relied on the AI assistant to provide deeper knowledge beyond what is directly present in the recordings.
Compared to frameworks on different levels of AI assistance \cite{mackeprang_discovering_2019}, the AI assistant \rv{AI assistant was simulated to behave in} the lower end of the proposed automation spectrum since the extent of its capability was to offer suggestions. 
\rv{In contrast to fully automated methods that are on the higher end of the automation spectrum \cite{mackeprang_discovering_2019},} our study showed that participants felt comfortable with this level of assistance and trusted the provided responses. 

\textbf{Product and task information} is important to usability analysis since UX evaluators need to identify when certain users deviate from the expected interactions.
In practice, UX evaluators may already be familiar with the product under test and are responsible for setting up the study tasks \cite{mixpanel_what_2022}. 
In this study, participants were given a list of the study tasks and a brief description of the product background before beginning their analysis. 
However, the finding that 16.6\% of total questions fell in this category suggests that participants required detailed information on the product and tasks for robust analysis.
Interestingly, this category accounted for 22.1\% of questions in the text condition but only 9.7\% in the voice condition. 
Since the same descriptions were given to all participants, further research is warranted to understand whether this difference was due to the interaction modality or variations in participants' analysis habits from the between-subjects design.

The \textbf{user demographics} category only made up 5.5\% of all questions, but we expect that UX evaluators may ask less of these questions in practice.
Since UX evaluators are often responsible for defining the inclusion criteria when setting up a usability test, they would have knowledge of the user demographics \cite{mixpanel_what_2022}. 
Questions in this category were more frequently asked in the text condition (7.7\%) than the voice condition (2.8\%). 
Considering that both the product and task information and user demographics categories were related to background information, perhaps participants assigned to the text condition were slightly more interested in contextualizing the results than those in the voice condition. 

\subsubsection{Interacting with Conversational Assistants} \label{sec:conversational}
In comparison to prior work that offered non-interactive visualizations of ML-driven features \cite{fan_vista_2020, soure_coux_2021}, our design offered a conversational interface so that UX evaluators could interact with the AI and receive timely information as needed.
Some participants even felt that the AI assistant acted a ``colleague'', which supports prior findings \cite{lakkaraju_rethinking_2022}.
They mentioned having more trust in \textbf{objective information} than subjective responses, which makes sense as AI-driven interfaces were perceived as more positive and credible when they exhibited machine-like properties such as objectivity \cite{sundar_main_2008}. 
Participants also described that they were satisfied with the answers provided by the assistant since it was straightforward and did not exhibit emotions. 
They appreciated the \textbf{professionalism of the AI assistant} since it was not pretending to be human (e.g., \textit{``I liked that the AI assistant didn't have a name or personality, that would feel unprofessional''} -P5-T). 
Prior work showed that dialogues imitating humans outside of the expressed purpose of a conversational agent could lead to negative experiences \cite{kim_understanding_2022}, thus conversational agents should keep their dialogues focused on the specific tasks that they were designed for. 
Although participants in this study liked that the AI assistant admitted knowledge gaps, responses that only contain \textit{``I don't know''} may lead to negative perceptions of its usability \cite{lee_i_2021}.
Thus, when it is difficult to provide a complete answer, conversational assistants may include reasonably related information (e.g., \textit{``I don't know why the user only clicked the menu, but I think it's probably because he didn't know he could scroll down.''}) as it demonstrates a greater effort to help \cite{lee_i_2021}. 
However, whether exhibiting higher effort to help would be appreciated in the context of usability analysis remains to be explored. 


\subsection{Text vs. Voice Interaction Modalities}

\subsubsection{Number of Questions}
Our results showed that questions in the text condition accounted for 56\% of all collected questions, and for the website video, participants using the text assistant asked significantly more questions than voice assistant.
One key difference that impacted the efficiency between the two modalities was the ability to edit questions since they were expected to be in complete sentences.
For text interactions, it was easy for participants to directly edit in the chatbox (e.g., deleting some words and reuse the rest of the typed text to complete the question.)
However, for voice interactions, participants needed to think about their questions before asking it out loud.
If they wanted to change the phrasing of their question, they would need to start from the beginning to say the whole question again.
Another difference was in the invocation of the AI assistant. Participants could directly type their questions to the text assistant, while they needed to say the wake phrase \textit{``Hey AI assistant''} or press and hold the microphone icon, which led to extra time.
Although a prior study showed no differences in the number of annotations made using speech or text for a collaborative writing task \cite{neuwirth_distributed_1994}, the conditions were different in that participants directly recorded or listened to voice annotations without the need for invocation nor transcription.
Another study reported higher dropout rates for participants in the voice condition than text due to receiving low-quality transcriptions \cite{kang_understanding_2017}.
\rv{In our study, participants did not drop out due to transcription errors since the moderator responded to questions as they heard them and provided an appropriate answer. 
Furthermore, transcription errors did not change the fact that participants asked a certain question in the first place, although it may have impacted the speed and perceived efficiency of the participants when they repeated the same question.}

\subsubsection{Length of Questions}
Although prior work found that spoken queries were longer and more conversational in movie recommendation systems \cite{kang_understanding_2017} and that spoken comments were longer in collaborative writing tasks \cite{neuwirth_distributed_1994}, our results showed no significant differences in the length of questions between speaking and typing.
Since longer messages tended to be positively associated with engagement \cite{chae_effects_2020, van_heerden_potential_2017}, our results suggest that text and voice assistants offered similar levels of engagement. 
Further investigation is needed to see if this observation is context-dependent.

In this study, participants using both text and voice assistants mentioned making an effort to use simple phrasing, which echoes prior research that people tended to use more restricted vocabulary with conversational agents than with humans, but they can easily adapt their language \cite{hill_real_2015}.
In the voice condition, participants also tried to speak more slowly and succinctly so that their questions could be accurately transcribed. 
Prior research with commercial voice assistants also showed that participants made use of tactics such as removing words other than keywords, reducing the number of words used, using more specific terms, altering enunciation, and speaking more slowly and clearly \cite{luger_like_2016}. 
Despite the differences with prior work on the modality where longer messages occurred \cite{neuwirth_distributed_1994}, we found that the content of the long messages were the same---they all consisted of explanations. 
Longer messages in the the collaborative writing task contained explanations of reviewers' suggestions \cite{neuwirth_distributed_1994}, whereas longer messages in this study contained explanations of why participants offered a certain design suggestion.

\subsection{Design Considerations}
Our study generates four design considerations to answer key questions for the future development of conversational AI assistants for UX evaluation. 

\subsubsection{What is the range of questions that future AI assistants should anticipate?}
\textbf{Future conversational AI assistants should anticipate questions about user actions, user mental model, help from AI assistant, product and task information, user demographics, as well as consolidated analysis from multiple recordings.}
When delving into the \rv{five} main categories, questions in the \textbf{other users' actions} subcategory were particularly interesting. 
We followed the design of prior work that analyzed individual usability test recordings \cite{fan_automatic_2020, fan_vista_2020, soure_coux_2021}, however, participants wanted the AI assistant to be more knowledgeable across multiple recordings.
Prior work showed that 66\% of UX evaluators considered the frequency of an issue across multiple participants when deciding the severity of a usability problem \cite{kuang_merging_2022}.
P7-T mentioned that \textit{``the assistant would be more useful if it had some meta knowledge from all the videos and could compare different videos to find common themes since that takes more time and effort, the average numbers are what we need to present in the final report of a usability test.''}
Based on the participants' responses, we suggest that future assistants would be more helpful if they had context across multiple users so participants could \textit{``get a sense of how typical or atypical certain user actions were''} -P2-T.
Thus, future research should explore how to connect analysis from multiple recordings of the same task and consolidate results into useful information for UX evaluators.

\subsubsection{How can automatic methods and human-AI collaborative methods be leveraged together in the future design of AI assistants?} 
\textbf{Provide both summary visualizations with objective information and conversational interfaces that could answer subjective questions as a means to seek explanation and clarification.}
We found that questions in the \textbf{user actions and mental model} category were typically objective, which meant they could be answered with little controversy.
Questions in the \textbf{help from AI assistant} category were typically subjective, which required judgment and may involve uncertainty. 
Compared to subjective responses, participants were more trusting of factual and objective information, which we identify as an opportunity for automation. 
Common questions in the \textbf{user actions and mental model} category included what users did, how users felt, and whether they found it difficult to complete a task. 
Existing algorithms for translating videos to natural language could also be applied to usability test recordings to describe user actions \cite{venugopalan_translating_2015, hendricks_localizing_2017}. 
Emotions can be detected either from facial expressions (e.g., \cite{chowdary_deep_2021, jain_extended_2019}) or speech of the user (e.g., \cite{issa_speech_2020, imani_survey_2019}). Recently, some commercial platforms like UXTesting incorporated emotion detection in their UX testing services \cite{uxtesting_uxtesting_2022}. 
Prior work on difficulty prediction relied on psycho-physiological sensors \cite{fritz_using_2014} which may not be feasible in remote usability tests, and navigational speed \cite{liu_predicting_2010} which may be subject to individual differences (e.g., young vs. older adult). 
Overall, future work should investigate how state-of-the-art research algorithms may generalize to usability test recordings instead of benchmarking datasets and verify the effectiveness of new ML-driven features in commercial tools. 

If algorithms were able to accurately extract user actions from recordings, participants mentioned that it would be helpful to see quantitative statistics common to all recordings (e.g., clicks, time per task) as a summary, which would reduce the need to repeat questions for every recording.
This feedback suggests that participants would accept these types of information to be automatically detected and presented on a dashboard.
However, the conversational interface offered many benefits as discussed in Sec \ref{sec:conversational}, \rv{which includes} making participants feel as if they were collaborating with a colleague and the chat window acting as a record of their analysis logic.
\rv{There are fundamental differences between non-interactive visualizations that present information regardless of evaluator needs and interactive conversational assistants that provide information on demand and more evaluator agency. 
The unique advantages} of both methods could be combined by providing UX evaluators with an overview of objective information in the form of a dashboard while having the opportunity to ask higher-level subjective questions using the conversational interface.
\rv{A comparison of the two approaches would also be an interesting future study, built upon the results presented in this study.}

\subsubsection{How should interaction modalities be designed for future conversational AI assistants?}
\textbf{Future designs should give UX evaluators the option to use either text or voice on demand since they provide unique advantages in different scenarios.}
We conducted a thorough comparison on the length and content of participants' questions, and the six subjective ratings of text and voice assistants.
Our study found that there were benefits and drawbacks to both, with text being rated as significantly more efficient. 
In addition to the inherent task constraints like the voice responses interrupting audio from the recording or text responses distracting attention from the video, the choice of text vs. voice is also impacted by the external environment. 
With the continuation of remote work, UX evaluators who are working at home may find the voice assistant more convenient due to the ability to multi-task as mentioned by multiple participants. 
However, in an office setting, text may be the main modality of interaction so that UX evaluators do not interrupt their coworkers in the near vicinity. 
\textit{``I think speech may be really convenient since I can just say my questions while typing notes on the usability problems, but I would only use it when I'm at home and not in the office with people around''} -P7-T. 
Thus, future designs should allow UX evaluators to switch between modalities since they provide unique advantages in different scenarios.

\subsubsection{How can conversational AI assistants be better introduced to UX evaluators?}
\textbf{For UX evaluators to make full use of the assistant, they should receive a comprehensive introduction to the AI assistants' capabilities prior to engaging in analysis.}
Participants mentioned that although they had experience using conversational assistants like Siri, this is the first time they interacted with one designed for UX analysis. 
Even though they were given a description of the AI assistant, this explanation did not provide sufficient detail and left them wondering about the extent of its capabilities, which led to questions in the \textbf{assistant capabilities} subcategory.
P1-T and P2-T mentioned that they would like an \textit{``introductory video that describes all the features of the assistant''} so that they would be familiar with the functionalities beforehand.
Demonstration videos, tutorials with a thorough walk-through of features, and practice sessions were proven effective in studies where participants used human-AI collaborative tools \cite{fan_human-ai_2022, soure_coux_2021, lai_human-ai_2022}.
Thus, future studies should provide a combination of resources that offer a comprehensive introduction to the AI assistants' capabilities prior to analysis. 

\subsection{Limitations and Future Work}
Our research took the first step to designing a conversational assistant to help UX evaluators with usability analysis. We used two short videos (one website and one app) \rv{that are not representative of all types of existing usability tests}. 
Since the number and types of questions asked might be affected by the length and content of the videos, future work should collect more usability test videos of different products and tasks to better understand whether these categories and their relative proportions remain consistent.
\rv{Testing more products would also improve the generalizability of the proposed method and validate if the conversational approach is effective across different products.}

\rv{We had a limited number of participants, which may have resulted in fewer collected questions and some non-significant differences in subjective ratings.
With more participants, we may determine whether the observed differences in the categories of questions asked between text vs voice conditions are still consistent. 
However, one study is likely insufficient even with more participants when determining if a trend holds \cite{hornbaek_is_2014}. 
Thus, future work is warranted to collect a wider range of questions from UX practitioners and affirm the observed differences between modalities.}

We used an existing speech-to-text library which did not always produce accurate transcriptions. This may have impacted the experience of non-native English-speaking participants who used the voice assistant. 
Since about 75\% of English speakers \rv{worldwide} are non-native, which includes many UX evaluators in industry \cite{stevens_viewpoint_2019}, there is a need for algorithms to become more inclusive of different accents. 
However, this limitation might not have severely affected our findings as \rv{only 2.1\% of transcribed words were misunderstood and} the moderator answered based on what they heard.

Due to the Wizard-of-Oz design, the moderator needed time to type in the responses once a question was asked. Thus, many participants felt that \textit{``as a tool, it was too slow.''}
Based on their prior interactions with conversational assistants, they expected that the responses would be instantaneous.
When designing AI assistants that are backed by algorithms, we expect the speed of responses to improve. 
However, other participants pointed out that they felt the AI assistant worked as a colleague, in which case it might be expected for them to take time before responding.
It remains an open question of whether it is preferable for the AI assistant to provide answers instantaneously like a robot or with some delay as a human-like colleague. 

\rv{The simulated AI assistant in this study had limited capability, which may have impacted the depth of questioning.
Participants may have follow-up questions based on the response from the AI assistant, which was stymied if the assistant could not provide an answer. 
We focused on delivering the fidelity of the experience since the AI was not real, and our results contain a reasonable set of questions and a baseline for perceived usefulness ($Md = 4$ for text and $Md=3$ for voice). 
Thus, we can extrapolate that any AI assistant beyond our representation ought to offer a deeper set of questions and higher perceived usefulness. 
Future work could explore interactions with a more powerful AI and add to the current dataset.
Furthermore, it is interesting to explore how UX evaluators' behavior (e.g., categories and frequency of questions asked) might change with the long-term usage of the AI assistant.}

In sum, we suggest future work that explores nuanced interactions between UX evaluators and AI assistants which have capabilities outlined by the collected dataset. Even though we conducted the study with UX evaluators, findings about interaction modalities and design considerations can be applied in future work for other domains where AI is involved in the decision-making process. 
\section{Conclusion}
The goal of this research was to understand how best to design a conversational assistant that helps UX evaluators by investigating \rv{(1)} what questions evaluators had while conducting analysis; \rv{(2)} how they wanted to ask these questions\rv{; and (3) whether they found text and voice assistants useful}. 
Based on a Wizard of Oz design probe with 20 participants, we found that participants were interested in \rv{five} categories of information: \rv{\textbf{user actions} (30.2\%), \textbf{user mental model} (21.5\%)}, \textbf{help from AI assistant} (26.2\%), \textbf{product and task information} (16.6\%), and \textbf{user demographics} (5.5\%).
Participants either adopted an \textbf{assistant suggestions last} or an \textbf{assistant suggestions first} strategy when interacting with the AI assistant. 
\rv{More questions were asked to the text assistant and} participants felt that it was significantly more efficient than the voice assistant. 
Based on these findings, we provided design considerations for future conversational assistants which include consolidating information across multiple usability test recordings, providing visualizations for objective information and conversational interfaces for subjective information, and allowing evaluators to choose the modality of interaction.
\rv{Future work is warranted to affirm the trends for the categories of questions found in this study and whether the trends would hold for long-term usage. Furthermore, it is interesting to explore the optimal timing of the responses from the AI assistant.}
In sum, our work has taken the first step to apply conversational assistants to the task of UX analysis and identified advantages and trade-offs between text and voice modalities.


\begin{acks}
This work is supported in part by the Natural Sciences and Engineering Research Council of Canada (NSERC) through the Discovery Grant.
We also thank all our participants for taking the time to share their feedback. 
\end{acks}

\bibliographystyle{ACM-Reference-Format}
\bibliography{main}


\end{document}